\documentstyle[12pt]{article}
\begin{document}
\vskip 2 cm
\begin{center}
\large{\bf
A " LAYERS OF REALITY TO A WEB OF INDUCTION " HYPOTHESIS}
\vskip 3 cm
{\bf Afsar Abbas}\\
Institute of Physics, Bhubaneswar - 751005, India\\
{afsar@iopb.res.in}
\vskip 4 cm
{\bf ABSTRACT}
\end{center}
\vskip 2 cm

It is shown that as knowledge is structured, it comes in 
modules. This provides different " layers of reality ". 
Each layer of reality has its own distinctive inductive logic 
which may differ from that of the others. 
All this is woven together to form a " web of induction "
in a multidimensional space. It is the overall resilience, 
firmness and consistent interconnectedness
of the whole web which justifies 
induction globally and which allows science to continue to
"read" nature using the inductive logic.

\newpage

Advanced knowledge of science and other disciplines is imparted
through Universities. Each university has departments and 
each department specializes in a particular discipline. Different
departments mostly work independently of each other. 
Most of the time they 
are not even aware as to what is being taught in the other
departments. They continue to do so without 
in anyway compromising 
their ability to impart knowledge in their own discipline. 
Hence this is the most straight forward acknowledgement of the
fact that scientific knowledge is
structured and comes in modules. By and large each module of
knowledge may exist independent of the other modules of knowledge.   

Succinctly we may state that scientific knowledge, 
as a mapping of reality,
is structured into "layers". In common usage the word layer stands
for two dimensional surfaces in a three dimensional world. 
Let us not confine ourselves to this three dimensional world here.
When talking of the dimensionality of any reality one necessarily
means the dimension ( or number of degrees of freedom ) of the
minimum number of variables needed to describe that particular 
aspect of nature completely. 
Hence here a layer would mean a surface
in this "large dimensional" field of reality. 
Hence the layers of reality that we are talking about has
multidimensionality built into it.

So, broadly speaking, scientific knowledge  
can be considered as not being an interrelated or
interdependent compendium
of facts and relationships. It is "almost" discrete and exists in
different compartments which are 'almost' independent of each
other. In fact it is
entirely because of this feature of nature that we have been able
to 'uncover' or understand 
nature, part by part and one at a time. 
In fact, had it been not so, then one may wonder as to how
scientists would have ever been able to 
acquire any knowledge whatsoever ?

As scientists, we have gone into different
layers one at a time. Once in a while one encounters a boundary
between two layers and that creates unanticipated complications
and difficulties. In tackling these, one often discovers other 
hidden layers of reality. 
However. if it turns out that a particular boundary is indeed
unbreakable, then that too gives 
information about still other aspects of what reality is
all about. Hence knowledge is not static. It keeps on changing and
evolving. 
 
In physical science, there is a neat way of
understanding these layers of reality. Simply stated, the 
different layers of reality in physics have to do
with different energy or length scales. Without going into
mathematical details, short length scales imply larger
energies in a particular physical phenomenon and larger length
scales mean smaller energies 
available for relevant physical processes. 
It has been a major achievement
of the physical sciences that it has uncovered the fact that
nature is built upon different length/energy scales. And that
these energy scales are a few and finite in number. 
As long as one is working
within a particular energy scale, one can by and large 
ignore effects arising from the other energy scales. 
Boundaries where the two energy scales meet require careful
handling as that may create unexpected complications.
A proper understanding of these has always  been found to be
extremely enlightening as well. 
However, as long as one stays away
from these complicating cases, then one has a 
good and independent sub-discipline of physics.

Gravitational interaction is dominant at large distance
scales. Coming down to smaller scale of a few Angstrom,
the gravitational interaction can be completely ignored
for all practical purposes and the
molecular forces start to manifest themselves. The
energies involved here are the ones available in ordinary
combustion of wood, paper etc. A little more energy is available
at say a little smaller distance of only one Angstrom or
so. At this level electrons are bound to protons to form atoms.
The relevant energies are approximately an  electron volt (in a
particular unit) or so. 
At this scale, one can study atomic physics without worrying 
about nuclear forces.
Nuclear forces start becoming significant as one goes to a much
smaller distance of a fermi or so.
The energies are now measured in a million electron volts or so. 
The whole discipline of chemistry need not worry about any other
scale than these two ( atomic and molecular ) to do its work!

One may treat these different energy scales as opening 
up of different layers of the "onion" of reality. In
nuclear physics ( which is relevant for nuclear power generation
and for nuclear bombs like the ones dropped on Hiroshima and
Nagasaki ) the relevant energy is about a million electron volts.
Still another layer of onion is known to exist at around 
a billion electron volt of energy.
At this scale quarks start manifesting themselves
explicitly in nuclei.
Still another scale occurs at a much higher scale of
energy  and is popularly called the Planck scale.
A lot of bizarre things are supposed to happen at the Planck scale
and today it forms a frontier of research in physics 
( Callender and Nuggett (2001) ).
Even the difficult case of the boundary problems, wherein
two scales meet, interesting things happen. For example  
phase transitions ( like 
ice melting of into water ) reveal still other basic features of
the mathematical reality - the significance of 
irrational numbers like the golden ratio etc. 

It seems that the reason that so far the 
philosophers of science have failed to appreciate the 
significance of the intrinsic existence of layers of reality 
as presented here is because 
intuitively they have been believing in the existence of a global,
uniform and all encompassing reality - 
which they have been trying to uncover, if not always through 
physical arguments than by using metaphysical arguments.

So far no such universal physical reality has manifested itself. 
The reality which has become clear in recent decades consists of
almost disjoint sets woven in a structured whole. 
It is nature which has forced the scientists to 
accept the fact that it is structured and consists of 
different layers of reality.
Once we have understood this fundamental structuring of 
the scientific knowledge of nature, 
then it will allow us to tackle the problem of
induction as well.

Since the time of Hume the logical method of induction has been
taking centrestage in the thoughts of mathematicians and
scientists in general and philosophers 
in particular. There has been
an onslaught against induction. It has been warding off these
attacks but it has had its back to the wall.  Basically
the only reason it has managed to survive so far 
is because the sciences, wherein it is extensively used, have
actually been "progressing". 
Whatever one may mean by the word progress, 
science and its associated technology is indeed changing
the world almost on a daily basis at present.

The arguments for and against induction are well documented
in text books ( Ladyman (2002) ) and other compilations
( Balashov and Rosenberg (2002) ). 
We do not intend to go into the detail of the same here.
however, still to put the issues in proper context
let me quote Broad (1887 - 1971) who very crisply called
induction, "the glory of science and the scandal of philosophy".

In terms of the layers of reality,
if one has a theory which explains the
reality at a particular layer, it had better be "complete" 
in as much as it would give a consistent description of
the physical reality manifested in that layer. This theory will
involve its empirical justifications, mathematical framework and 
reliable predictability. Hence a particular induction shall
be applicable in that regime. In simple terms, one has to agree
to this induction as it actually "works". 
It works because in the regime 
under discussion, these set of physical and mathematical arguments
of the relevant inductive framework explain 'all' empirical
information and make predictions which are found to be correct.

It is a common feature, that scientists working in a particular
discipline ( describing a particular layer of reality ) would
soon start finding limits to the applicability of 
that particular theory.
They would find that there are situations wherein the particular
inductive logic inherent in the description of that layer of
reality actually fails. Hence the scientists are forced to define
boundaries ( in terms of some physical parameters like say high
energy/low energy or small distance/large distance or 
low temperature/high temperature etc.) 
within which a particular inductive logic works and beyond 
which it fails. 
 
But this does not mean that the particular theory describing the
reality manifesting itself within that particular layer
is wrong. It was correct in as far as it was applicable -
empirical aspects incorporated in the theory and useful
applications ( if there be any ), 
as technological spin offs arising from the particular theory, 
would testify to it. 
But it is not universally and globally correct for all the 
situations. In struggling with these limitations of a particular
theoretical reality, one does further experimentations with
different mathematical models, until one finds that
he/she has reasonable understanding of another layer of reality.
This forces one to appreciate and use 
another set of inductive logic and so on. This is 
the way that nature has been unfolding itself to the scientists.

As an example Newton's gravitational mechanics was useful to
explain Kepler's laws of planetary motion. It was a consistent
theoretical framework which was very successful. Then came along 
Einstein's theory of special relativity in 1905 and which shook
the foundations of classical theory of gravity. Did it prove
Newtons gravitational force laws wrong? If one reads the
books in philosophy of science - in general the answer is that
yes, Einstein's theory proved Newton's theory wrong. 
It is unfortunate that many philosophers of science actually
think that Einstein's theory proved Newton's Theory wrong. This
involves blatant misunderstanding of the physics involved. 
In fact, even today 
one can use Newtons laws of motion with great confidence to
understand planetary motions up to a level of accuracy which is
acceptable in most of the situations. 
Only when one requires an accuracy to a much higher 
place in decimals and when the velocities of objects involved are
much higher (ie approaching the velocity of light) does one 
need to incorporate the corrections arising from Einstein's
theory. In reality Einstein's theory defines the limit (in terms
of relative velocity etc.) on the Newton's Law of Gravitation. One
should bear in mind that when a physical law has been empirically
verified to work under certain conditions (like temperature,
pressure, distance, energy etc) then 
under the same conditions it will continue to work the
same way always. If at all there arises a situation that one
encounters a failure of the same, then it just shows that one's
initial understanding was limited as it did not take into account
the situation in which it failed. 
 
Another example, wherein one would explore 
different layers of reality
in terms of different inductive/theoretical framework, is that of 
classical thermodynamics description 
going over to kinetic theory of gases description and
thereafter to
statistical mechanics description 
( which itself may go over from classical
statistical mechanics to quantum statistical mechanics both for
fermions and bosons). 
At each layer of reality, there were experimental and empirical
statements which were translated into a suitable theoretical
framework with its relevant concepts and proper mathematical
language. Predictions were made and confirmed. One gained
confidence in one's inductive logic when the same was used to
'control' reality by making relevant innovative technology 
to serve mankind and which could not
have been visualized without that particular induction involved. 
Then as one gains control over various physical parameters like
temperature, pressure and density etc one may be forced to go to
another layer of 
reality with its own inductive logic. Today quantum statistical
mechanics is trying to extend and establish its own limits of
applicability.

Still another example of layers of reality each with its own
inductive logic is that of 
geometrical optics going over to electrodynamics
and thereafter to quantum optics. All inductive logic in each 
subdiscipline  is accurate and reliable within its own limits. It 
is never wrong or inapplicable in these regimes. And each
furnished its own technology which was applicable within the
limits specified for that particular layer of reality and its
corresponding inductive logic.   

Understanding the boundaries between different layers is a more
challenging proposition. 
For a few such cases one may understand it by utilizing the  
limiting process 
( for example as in the mathematical language of calculus ).
Hence one finds that
in the limit of low density and high temperatures, quantum
statistical mechanical description would go over 
into the classical statistical mechanical one. 
In other cases like those of phase transitions, like as
in the early universe, as to how matter particles gained masses, 
exotic mechanisms like the Higgs mechanism
in the Standard Model of particle physics are invoked.
At present these boundary
situations between different layers of reality are under active
investigation and one looks forward to a deeper understanding of
nature arising from such endeavours.

Note that induction in "linear" within each layer of reality. By
'linearity' here I mean that inductive correlations are understood
in a direct manner in terms of a straightforward interpretation of
theoretical terms with the experimental reality. 
When one jumps to the next layer of reality
then another set of inductive laws are found to be
applicable. Those are 'linear'  themselves. But what about
the boundary of the two layers of reality? I suggest that this
boundary involves a " non-linear " jump. One goes from one
"linear" inductive logic of one layer of reality to another set of
independent "linear" inductive logic of the other layer through a
" non-linear " jump. 

As an example of this non-linear jump let us look at the 
early universe scenario in particle physics. As we stated a
little earlier, the masses of particles like electron, quarks etc
arise through a mathematical technique called 
the Higgs Mechanism. This is a non-linear process
with complicated mathematics.
As such this defines the boundary between two layers of 
reality which themselves are linear. 
Above the Higgs mechanism scale, the electrons are massless
and are understood in terms of a particular inductive logic and
below it another inductive logic manifests itself in 
terms of the Standard Model of Particle Physics. 
So a "non-linear" reality acts as a "knot" to
join together two different "linear realities".

I therefore propose " a web of induction hypothesis ".
In terms of induction one may define existence of a "web" of
induction. Different inductive logical systems are 
correlated with each
other through limits to form a web of induction. This web of
induction is what justifies induction. 
There is no universal and global inductive logic.
Induction is justified because different layers of reality with
their own limited but justified inductive logics hang together in
the form of a "web". 
The linear part of the induction ( at a particular layer of
reality ) forms the "thread" part of the web. 
The non-linear part of
induction ( for the boundary cases ) forms the "knot"
part of the web,
So knots in the web connect the threads together.
Induction is "correct" because the
resilience of the web of induction gives it strength and validity.

It is because different inductive
logics (which describe different layers of reality) from a
consistent and solid web in a multidimentional space 
is that one may use these to describe the reality of nature.
Each thread of the web can only be understood in its total
involvement with the rest of the threads as well as knots. You
cannot cut it at a point and hope that the rest will remain
intact. It is this web of induction 
that gives consistency to the whole
scientific enterprise. 

Note that "induction" gets justified as a valid means of
acquiring knowledge not because induction itself forms an 
essentially one single global entity - as has implicitly been 
assumed by most of the philosophers of science. 
As shown here "Induction", gets
a global justification because it consists of 
several interconnected parts which are  
relevant for a consistent description of 
the different layers of reality.
The whole thing stands firm, resilient and consistently 
interconnected in the form of a web.


\vskip 2 cm
{\bf REFERENCES}
\vskip 1 cm

Balashov, Y. and Rosenberg, A. (Ed.) (2002), " Philosophy of
Science - contemporary readings ", Routledge, London

Callender, C. and Nuggett, N. (2001), " Physics meets philosophy 
at the Planck Scale ", Cambridge University Press, Cambridge

Ladyman, J. (2002), " Understanding philosophy of science ",
Routledge, London

\end{document}